\documentstyle[draft]{mn}

\def\lapp{\ifmmode\stackrel{<}{_{\sim}}\else$\stackrel{<}{_{\sim}}$\fi}
\def\eqsim{\ifmmode\stackrel{\sim}{{=}}\else$\stackrel{\sim}{{=}}$\fi}
\def\gapp{\ifmmode\stackrel{>}{_{\sim}}\else$\stackrel{>}{_{\sim}}$\fi}

\title[Emission altitudes  in young and old radio pulsars]
{Emission altitudes in young and old radio pulsars}

\author[J. Kijak ]
{Jaros\l aw Kijak \\
~~\\
J. Kepler Astronomical  Center, Pedagogical University, Lubuska 2, 
PL-65-265 Zielona G\'ora, Poland; jkijak@astro.ca.wsp.zgora.pl}

\begin{document}

\maketitle

\begin{abstract}
The paper presents
a comparison of emission altitudes in very young and very old radio pulsars.
The author  confirmes that  altitudes at which radio emission
at a given frequency is generated depend on pulsar period  and on
pulsar age, although the latter dependence is quite weak.
\end{abstract}


\section{Introduction}
The radial location of the radio emission regions in  pulsar
magnetospheres was discussed by several authors. Using  various methods
reviewed by Cordes (1992, 1993),
constraints on the emission radii were obtained in a number of papers. 
These estimates show
that the pulsar emission region is relatively compact and lies near
the surface of the neutron star, apparently closer than about 10 per cent
of the light-cylinder radius (Cordes 1978; Blaskiewicz, Cordes 
\& Wasserman 1991; Phillips 1992; Gil \& Kijak 1993; 
Hoensbroech \& Xilouris 1997; Kijak \& Gil 1997, 1998; Kramer et al. 1997;
Gupta, Bhat \& Rao 1999).

Assuming that pulsar is a rotating magnetic dipole we can 
calculate its  characteristic age $\tau=P/2\dot P$ 
(Manchester \& Taylor 1977), which is a good estimate of the true age 
providing that $P\gg P_0$, where $P$ and $P_0$ are current and initial spin
periods, respectively. 
Most pulsars are $10^{6} - 10^7$ years old, however,  one separates
young and old objects with characteristic ages $\tau$ about $10^4$  yr and
$10^9$ yr, respectively. It is important to note that,
the group of old pulsars contains
not only millisecond pulsars, but
also  pulsars with typical period $P\sim 0.5$ s (e.g. PSR B1848+04, B1952+29)
as well as recently discovered very old  pulsar J2144$-$3933 with the longest 
measured preiod 8.5 seconds (Young, Manchester, Johnston 1999). 
It is commonly believed that pulsars form two dinstinct populations:
high-magnetic field objects that are radio pulsars for relatively
short times and recycled objects which can be very old.

The millisecond pulsars (MSPs) have different evolutionary history compared
to normal pulsars. They are recycled pulsars, i.e. they acquired
their rapid rotation rates ($\sim$ a few ms) due to accretion-driven spin-up 
(Bhattacharya 1996). Consequently, their magnetic dipole 
field strength at the surface
$B_d=3.2 \times 10^{19}(P\dot P)^{1/2}$ G (Manchester \& Taylor 1977)
is relatively low ($\sim 10^8-10^9$~G), as compared with the whole population
of radio pulsars ($\sim 10^{12}$~G).
Previously, it was thought that MSPs also have  different characteristics
of radio emission from those of
normal pulsars. However, it has been recently argued 
(Kramer et al. 1998, 1999; Gil \& Sendyk 2000) that MSPs emission properties 
do not differ from typical pulsars.

\begin{figure*}[t]
\setlength{\unitlength}{1cm}
\begin{picture}(1,14)
\put(-10,-1){\includegraphics{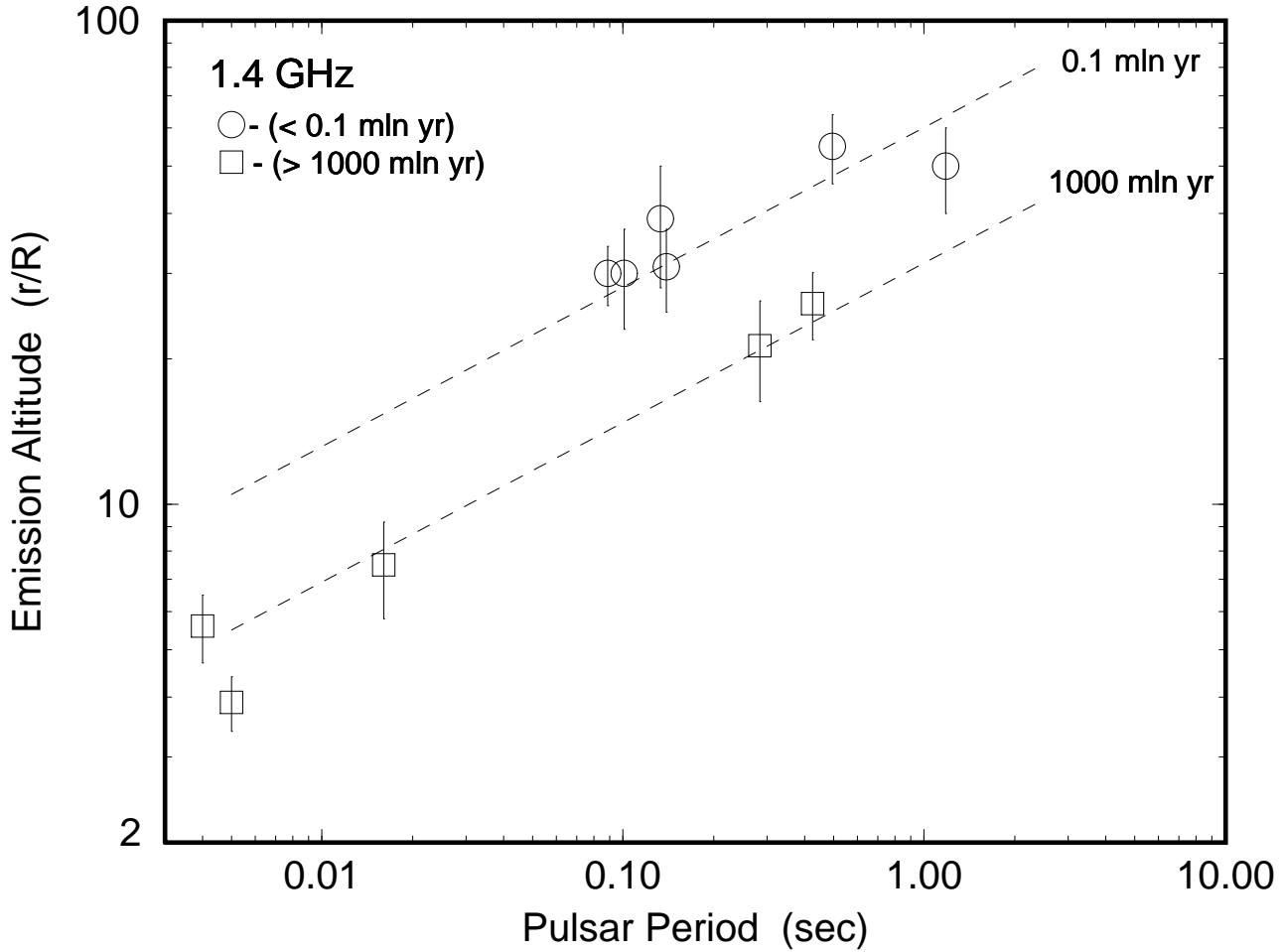}}
\end{picture}
\caption{
Emission altitudes at 1.4 GHz versus pulsar period for young (circles) and
old (squars) pulsars (data taken from Table~2 and for old PSRs: B1855+09, 
J1022+1001 and J1713+0747 from Paper~II). 
Formal weighted fit for pulsars
with $\tau_6 < 0.1$ gives $r_{\rm em}= R (55\pm 8) P^{0.25\pm 0.08}$
and for $\tau_6 > 1000$ gives $r_{\rm em}= R (35\pm 6) P^{0.38\pm 0.04}$.}
\end{figure*}

The semi-empirical relationship proposed by Kijak \& Gil (1997)
describe the dependence of emission altitude on pulsar period $P$
(in seconds), observing frequency $\nu$ (in GHz)
and characteristic age $\tau$ (in $10^6$ yr)
\begin{equation}
r_{\rm KG} = (55\pm 5)R \nu_{\rm GHz}^{-0.21\pm 0.07} \tau_{6}^{-0.07\pm 
0.03} P^{0.33\pm 0.05} ,
\end{equation}
where $R=10^6$cm is the neutron star radius. 
This was obtained from analysis of emission altitudes for a number of
pulsars at two frequencies. 
Emission altitudes were estimated using the pulse width method 
(see Section~2) based on a number of straightforward assumptions:
(i) the pulsar radiation is narrow-band with a radius-to-frequency mapping
operating in the emission region, (ii) pulsar emission is 
relativistically beamed tangentially 
to  dipolar magnetic field lines and (iii) the extreme profile wings 
originate at or near the last open field lines.
This, of course, involves an implicit assumption that the magnetic
field in the radio emission region is purely dipolar (see Section 3 for 
discussion).
The uncertainties given for various quantities in equation (1) follow from
systematic errors  in the pulse width $W$ measurements and random errors  in
estimations of the inclination angle $\alpha$ between the
rotation and magnetic axes and the impact angle $\beta$ of the closest 
approach of the observer to the magnetic axis.

First suggestion that radio emission altitudes at a given frequency
depend on the pulsar period ($r_{\rm em}\propto P^{\sim 0.5}$) was published 
by Gil \& Kijak (1993). 
This period dependence was further carefuly investigated  by 
Kijak \& Gil (1997, 1998; hereafter Papers I and II) and more precise
 relationship $r_{\rm em}(P)\propto P^{\sim 0.33}$ was obtained. 
Anyway, the emission region is apparently located close to the neutron star 
in short-period pulsars and further away in longer-period pulsars.
Alternatively, the ratio
$r_{\rm KG}/r_{\rm LC}$ decreases with increasing period (where $r_{\rm LC}$
is the light cylinder radius), what was quantitatively  demonstrated
in Papers I (Fig. 3b) and  II (Fig. 4a and 4b).
The equation~(1) 
describes  also  a radius-to-frequency
mapping (RFM) $r(\nu)\propto \nu^{-p}$ (where $p \sim 0.21$), 
which means that different frequencies are generated at different altitudes 
above the magnetic polar cap.
The RFM models of  pulsar emission region provide an attractive
explanation of why profile widths (usually) increase with decreasing frequency.
Kijak \& Gil (1998) studied the concept of RFM
using the multifrequency data of low-intensity pulse widths for normal and
millisecond pulsars. 
They obtained RFM  with the average exponent $\bar p = 0.26\pm 0.09$ 
(for 16 pulsars),  consistent with previous estimates
using different methods (see Table~4 in Paper II). 
They ultimately confirmed that the emission altitude depends
on frequency $\nu$, pulsar period $P$ as well as on the pulsar age $\tau$,
although the latter dependence was weak.
In this paper we  investigate a putative weak 
dependence of emission altitudes on
the characteristic age   indicated in earlier work
(equation~1), by selecting a sample of
very young and very old radio pulsars and applying to them the same 
analysis methods as used in Papers I and II.

\begin{table*}[t]
\begin{center}
\caption[]
{Emission altitudes for old and young pulsars at 1.4 GHz. 
We calculated $r_{\rm em}$ from Eq.~2 and $r_{\rm KG}$ from Eq.~1. 
Pulse width measurements $W_0$ 
at low-intensity level (i.e. $\approx 0.1$ per cent)
and calculations of opening angle $\rho_0$ are presented. 
References are marked for $\alpha$ and $\beta$ angles (see Paper~I).}
\begin{tabular}{cccccccc}
\hline
PSR  &  $\tau_6$ & $P$ &  $r_{\rm em}$ & $r_{\rm KG}$ &  $W_0$ & $\rho_0$ 
& Ref. \\
  &  ($10^6$ yr)& (sec) &  ($r/R$)      &  ($r/R$)    & (deg)   & (deg)  &  \\
\hline
B0833$-$45& 0.011 & 0.089 & 30$\pm 3$   & 32   &  39.0$\pm 1.4$ & $22.8\pm 1.3$
& R93 \\
B1800$-$21& 0.016 & 0.133 & 39$\pm 11$  & 35   & 137.$\pm 8.0$  & $21.2\pm 4.1$
& G94 \\
B1823$-$13& 0.022 & 0.101 & 30$\pm 7$   & 31   & 120.$\pm 4.0$  & $21.3\pm 3.6$
& G94 \\
B1727$-$33& 0.025 & 0.139 & 31$\pm 6$   & 34   & 86.0$\pm 7.0$  & $18.5\pm 2.7$
& G94 \\
B2334+61  & 0.040 & 0.495 & 59$\pm 7$   & 51   & 34.8$\pm 2.6$  & $13.5\pm 1.3$
& G94  \\
B1916+14  & 0.089 & 1.180 & 50$\pm 10$  & 68   &  16.2$\pm 2.5$ &  $8.0\pm 1.2$
& G94 \\ 
B1848+04  & 2818 & 0.284 & 21$\pm 5$   & 19   &  107.$\pm 9.0$ &  $10.7\pm 2.1$
& R93 \\
B1952+29  & 4168 & 0.426 & 26$\pm 6$   & 17   &  35.0$\pm 3.0$ &  $9.7\pm 1.6$
& LM88 \\
\hline
\end{tabular}
\end{center}
\end{table*}

\section{Pulsar characteristic age and emission altitude}

In this section we calculate emission altitudes for very young 
and very old pulsars  using 1.4 GHz data from Effelsberg Observatory.
In addition, data for two southern pulsars  B1727$-$33 (Gould \& Lyne 1998)
and  B0833$-$45 (Johnston, Nicastro \& Koribalski 1998) were included
because of their  young age.
All these pulsars (Table~1 and Fig.~1) 
were not analysed in previous work (Papers I and II).
For estimations of the emission altitude we use the geometrical
method (pulse profile width $W$ measurements at low-intensity level 
of the maximum intensity) described in Papers I and II. 

Assuming that the magnetic field in the emission region has a dipolar form,
the emission altitude $r_{\rm em}$ (at which radiation at a given frequency 
$\nu$ is generated) can be expressed in the form
\begin{equation}
\frac{r_{\rm em}}{R} = \left(\frac{\rho}{1^{\circ}.24\cdot s}\right)^2 P~~,
\end{equation}
where $\rho$ is the opening angle [in degrees] between
the magnetic axis and the tangent to dipolar magnetic field lines at points
where the emission corresponding to the apparent pulse-width originates,
and  $s$ is the mapping parameter 
($0\leq s \leq 1$) describing
the locus of field lines  on the polar cap ($s=0$ at the pole and $s=1$ at
the edge of the Goldrech-Julian (1969) circular polar cap). 
Opening angles 
$\rho=\rho(W;\alpha,\beta)$ are calculated
using the knowledge on the viewing geometry
(represented by  the inclination angle $\alpha$ between the rotation and 
the magnetic axes and  the impact angle $\beta$ of the closest 
approach of the observer to the magnetic axis;
see Paper I and references therein for more detailed explanation),
as well as  measurements of the pulse width $W$. 
Estimates of emission altitude (for $s\approx 1$) from equation~(2) 
are listed in Table~1.

\begin{table}[b]
\begin{center}
\caption{List of 37 pulsrs which are analysed in
Figure~2 and 3. The corresponding emission altitudes presented in Fig. 2 and 3
are taken from Paper I and II, as well as
from the Table 1 in this paper.}
\begin{tabular}{cccc}
\hline
B0329+54   & B0355+54 & B0450$-$18 & B0525+21  \\
B0540+23   & B0740$-$28 & B0823+26 & B0833$-$45 \\
B0834+06   & B0919+06   & J1022+1001 & B1133+16  \\
B1237+25   & B1642$-$03 & B1706$-$16 &J1713+0749 \\ 
B1727$-$33 & B1749$-$28 & B1800$-$21 & B1804$-$08 \\
B1823$-$13 & B1845$-$01 & B1848+04   & B1855+09   \\
B1859+03   & B1913+16   & B1915+13   & B1916+14   \\
B1924+16   & B1933+16   & B1952+29   & B2020+28   \\
B2021+51   & B2310+42   & B2319+60   & B2324+60   \\
B2334+61   &            &            &        \\
\hline  
\end{tabular}
\end{center}
\end{table}

In Figure 1, we present a plot of emission altitude $r_{\rm em}$
versus period $P$ for very young and very old pulsars corresponding to
pulse width measurements at~1.4 GHz. 
The data have been taken from Table 1 and from the Paper~II
(for MSPs: J1022+1001, J1713+0747 and B1855+09).  
It is noticeable  that emission altitudes differ slightly between very young 
and very old pulsars. A formal weighted fits to very young and 
very old pulsars yield 
$r_{\rm em}= R (55\pm 8)P^{-0.25\pm 0.08}$ 
and  $r_{\rm em}= R (35\pm 6)P^{-0.38\pm 0.04}$, respectively. 
We can now gather all estimates of emission altitudes from Papers I, II
and from this paper (Table~1)  to obtain a plot of emission altitude versus 
characteristic age for 37 pulsars (see Fig.~2 and Table~2). 
The periods of normal pulsars were divided into six groups:
(0.05~s$-$ 0.2~s), (0.2~s $-$ 0.4~s), (0.4~s $-$ 0.6~s), (0.6~s $-$ 0.8~s),
(0.8~s $-$ 2.0~s), (2.0~s $-$ 4.0~s) and data belonging to each group
were represented by different size of the circle (with biggest circles
corresponding to longest periods in our sample). 
This is consistent with equation~(1), showing that for a given characteristic
age, pulsars with longer periods emit their radio emission from higher
altitudes than those with shorter periods.
A formal weighted fit to all data points in Fig.~2 is
$r_{\rm em}= R (33\pm 1) \tau_6^{-0.07\pm 0.02}$, confirming a putative
weak dependence  $r_{\rm em}$ on pulsar age $\tau=\tau_6\cdot 10^6$
years (equation~1).

\section{Discussion}

In Papers I and II the radius-to-frequency mapping
and the period dependence of emission altitude have been  discussed 
in details. Here,
a detailed analysis of age dependence on emission altitude is carried out.
Results presented in Figs. 1 and 2 indicate that emisson
altitudes depend on the  characteristic age 
$r_{\rm em}(\tau_6)\propto  \tau_6^{-0.07}$,
which is consistent with equation~(1). In this analysis we used 37 pulsars 
at 1.4 GHz for which  all necessary information were available,
that is: very good quality 
profiles (pulse width measurements) as well as reliable estimates of 
$\alpha$ and $\beta$ angles.

\begin{figure*}[t]
\setlength{\unitlength}{1cm}
\begin{picture}(1,14)
\put(-10,-1){\includegraphics{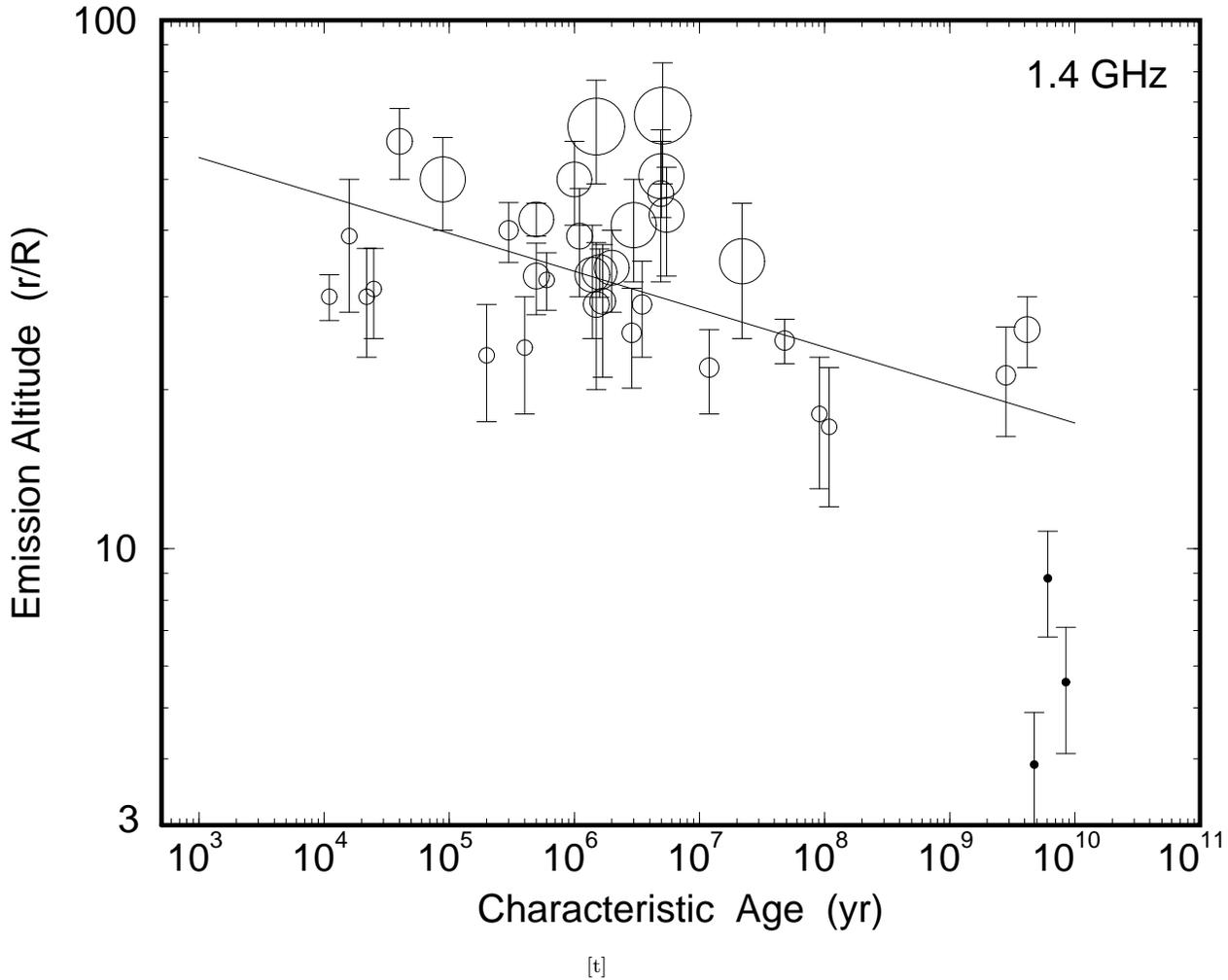}}
\end{picture}
\caption{
Emission altitudes versus characteristic age $\tau$ for 37 pulsars. 
The size of circles represents increasing period from 0.1~s to 4.0~s (see text
for explanation).
The filled dots correspond to millisecond pulsars.
A formal weighted fit gives 
$r_{\rm em}= R (33\pm 1) \tau_6^{-0.07\pm 0.02}$.}
\end{figure*}

It is important to notice
that the radio emission region for a group of very old pulsars 
(containing  millisecond and normal pulsars)
is located at correspondingly  lower altitudes than that of very young objects 
(see Figs.~1 and 2).
Three points in Fig.~2 deviate significantly from the rest of distribution.
These points correspond to 3 millisecond  pulsars in our sample.
One could think that this is an effect of 
nondipolar magnetic  field in the radio emission region of millisecond pulsars.
However, we do not regard these outlier points as strong
indication of deviation from
dipolar fields. In fact, the emission altitude at a given frequency is 
determined more strongly by the value of pulsar period than by 
its characteristic
age (see equation~1). Therefore, the emisson altitude of these milisecond
pulsars is about one order of magnitude smaller than in typical pulsars,
with periods in the range $0.1~-~4.0$ seconds (thus covering one order of
magnitude), much longer than the millisecond ones.
If one omits  three points corresponding to 
the millisecond pulsars,  then a formal weighted
fit is $r_{\rm em}= R(35\pm 1)\tau_6^{-0.04\pm 02}$, again consistent with
equation (1).

The surface magnetic field is estimated using the assumption that the main 
rotational energy loss is in the form of electromagnetic dipole radiation.
With a few further assumptions about the pulsar spindown mechanism
(Manchester \& Taylor 1977; Camilo, Thorsett \& Kulkarni 1994),
the surface magnetic dipole field component of a pulsar (inferred from its 
values of $P$ and $\dot P$) can be expressed in terms of pulsar
characteristic age $\tau = 0.5 \cdot P/\dot P$~[s] as follows
\begin{equation}
B_{\rm d} = 4 \times 10^{12} \tau_6^{-1/2}P~~~{\rm [G],}
\end{equation}
where $\tau_6=\tau_{\rm years}/10^6$.
If the magnetic field in the emission region is dipolar, one can
calculate its values at the emission altitude $r_{\rm em}$ as
\begin{equation}
B_{\rm r} = B_{\rm d} \left(\frac{r_{\rm em}}{R}\right)^{-3}~~~.
\end{equation}
From equations (1), (3) and (4),
one obtains 
\begin{equation}
B_{\rm r} \approx 2 \times 10^{7}\nu_{\rm GHz}^{0.63}\tau_6^{-0.29}~~~{\rm 
[G]}.
\end{equation}
The above equation  indicates that the magnetic field in the emission region
may depend on the characteristic pulsar age.
In typical pulsars with $\tau_6\sim 1$, the magnetic field in the emission 
region corresponding to 1.4 GHz should be about $10^7$~G.

Figure 3 presents a plot of emission altitude versus
the dipolar component of surface magnetic field (calculated from equation 3)
and the magnetic field in the emission region (calculated from equation 4).  
The magnetic field in the emission region at 1.4 GHz for normal pulsars 
($10^6 - 10^7$ yr) has values of about $10^7 - 10^8$ G.
Gil \& Kijak (1993) suggested that the magnetic field in the emission region
$B_{\rm r}$
is almost independent of the pulsar period and characteristic age.
This analysis clearly shows that very old pulsars (filled circles) and very young pulsars (circles with stars) have quite different magnetic field in
the emission region $B_{\rm r}$ (Fig. 3 and equation 5).

\begin{figure*}[t]
\setlength{\unitlength}{1cm}
\begin{picture}(1,14)
\put(-10,-1){\includegraphics{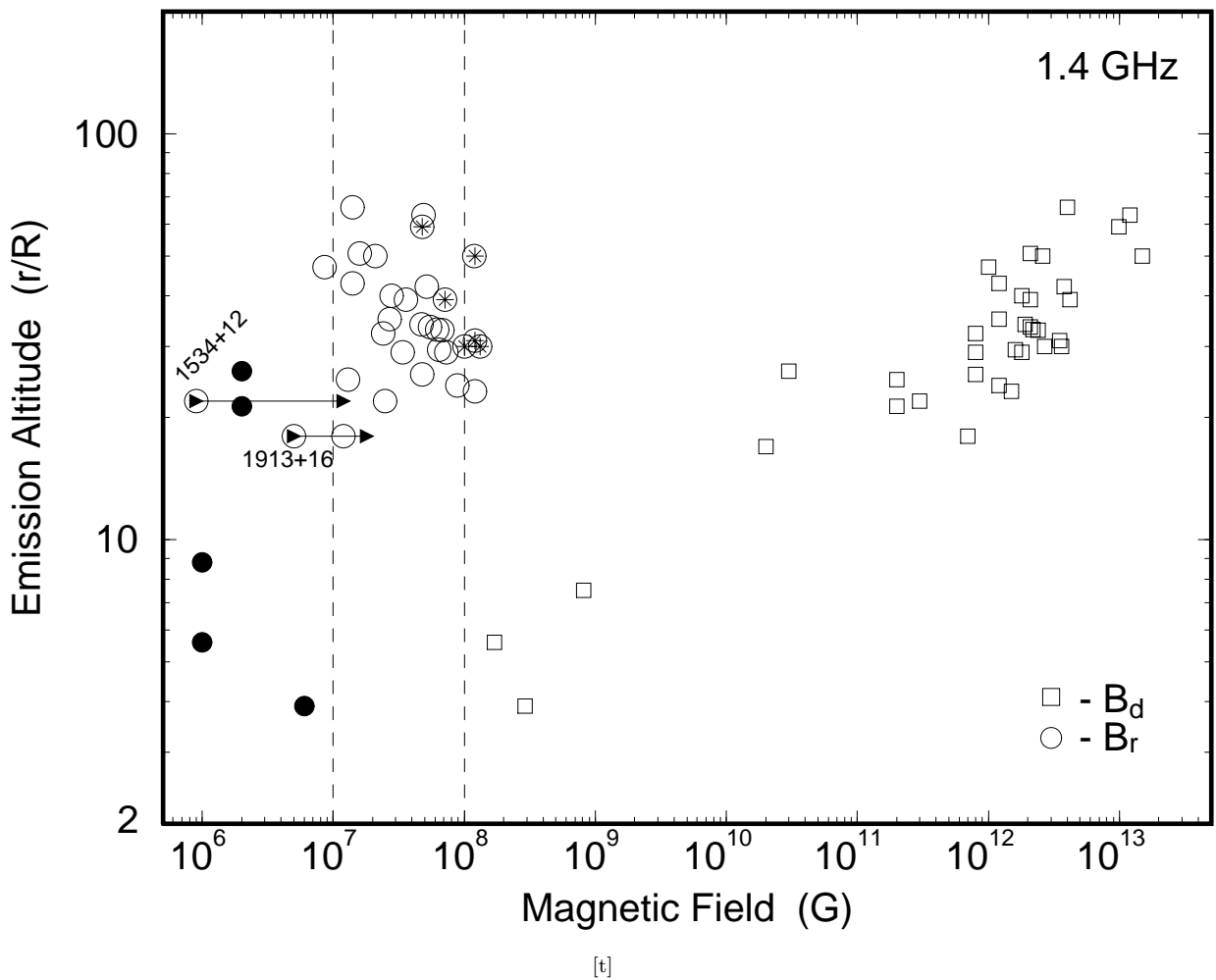}}
\end{picture}
\caption{Plot of the emission altitude $r_{\rm em}$ versus
two kinds of magnetic field, the surface magnetic dipolar field component 
$B_{\rm d}$ (squares)
and the field strength at the estimated emission altitude $B_{\rm r}$ 
(circles). 
The filled circles correspond to very old pulsars. Circles with star 
correspond to very young pulsars. The PSRs B1534+12 and B1913+16 are marked by 
the circles with horizontal arrows (see Section 3).}
\end{figure*}

Recently, a relationship between the characteristic and
true age was discussed in the literature. Several authors
(Lyne, Anderson \& Salter 1982; Camilo, Thorsett \& Kulkarni 1994; 
Bhattacharya  1996; Lorimer 1996, Cordes \& Chernoff 1998)
suggest that the true age is somewhat smaller than the characteristic
age computed from $P$ and $\dot P$ measurements. 
Arzoumanian, Cordes \& Wasserman (1999, hereafter ACW) argued that 
the three typical pulsars PSRs B1534+12, B1913+16 and B2127+11C are probable
younger than their characteristic age.
Let us consider two of them, PSRs B1534+12 and B1913+16,
for which the values of $\alpha$ and $\beta$ angles are known 
(Arzoumanian et al. 1996; R93, respectively).
In Table~3,
the geometrical emission altitudes $r_{\rm em}(W; \alpha, \beta)$, 
as well as semi-empirical altitudes $r_{\rm KG}$ calculated
from equation~(1) for characteristic
age (for $P$, $\dot P$) and ``true age'' (taken from ACW), respectively,
are listed.
One can see that $r_{\rm KG}$[$\tau_6$(ACW)] are closer to $r_{\rm em}$ than
$r_{\rm KG}$[$\tau_6(P, \dot P)$]. Moreover, as one can notice from Fig. 3, 
these two pulsars seem to have $B_{\rm r}$ values considerable lower than
the rest of the group of normal pulsar (between two dashed vertical lines).
It is worthy  of 
noticing, that even their characteristic age (not to mention the
``true age'') is much smaller than rest of the old pulsars (filled circles).
However, if we use ``true age'' (after ACW) in equation (5) then the verified 
values of $B_{\rm r}$ increase such that they fall into a region between
dashed vertical lines (which is marked by horizontal arrows). This indicates 
that the emission altitudes $r_{\rm em}$
and hence the magnetic field $B_{\rm r}$ in the emission region depends
on the true age rather than on the characteristic age. Of course, in most
cases, when the actual period $P$ is much larger than the initial period,
these values are the same.

\begin{table*}[t]
\begin{center}
\caption{Emission altitudes for two PSRs: B1534+12 and B1913+16 at 1.4 GHz.
Data for  the geometrical analysis  are taken from Paper I (B1913+16)
and Arzoumanian, Phillips, Taylor \& Wolszczan (1996) for B1534+12.
Emission altitudes of $r_{\rm KG}$ are calculated for characteristic age
and age proposed by Arzoumanian et al. (1999).}
\begin{tabular}{cccc}
\hline
PSR  & Geometrical  & Characteristic age & ``True age'' \\
     & analysis     &  analysis          &  analysis    \\
\hline 
B1534+12 & $r_{\rm em}=22R$ & $\tau_6=252$,~ $r_{\rm KG}(252)=12R$ & 
$\tau_6\le10$,~ $r_{\rm KG}(10) \ge 15R$ \\
B1913+16 & $r_{\rm em}=18R$ & $\tau_6=109$,~ $r_{\rm KG}(109)=14R$ & 
$\tau_6\le 5$,~ $r_{\rm KG}(5) \ge 18R$ \\
\hline
\end{tabular}
\end{center}
\end{table*}

Finally, let us discuss a possibility that the magnetic field lines in
the emission region are not purely dipolar.
The observational opening angle $\rho=\rho(W;\alpha,\beta)$ 
is calculated using geometrical
equation proposed by Gil, Gronkowski \& Rudnicki (1984), which
holds on the pulsar centered 
celestial sphere,  independently
of geometry of magnetic field lines in the emission region. 
This observationally inferred opening
angle $\rho=\rho_{\rm obs}(W;\alpha,\beta)$ is also the angle 
between magnetic axis and the tangent to  magnetic field lines 
at points where
the emission correspondig to the apparent pulse width $W$ originates.
For dipolar field lines $\rho$ is described by equation~(2),
with $s=1$ for last open field lines corresponding to low level 
emission below 1\% of the maximum intensity (e.g. Table~1).
For a simple model of star-centered quadrupole field in the radio emission
region, the opening angles 
$\rho_{\rm q}= 0.8 s (r_{\rm q}/R)P^{-1/2}$ deg (where $s<1$), as compared
with $\rho_{\rm d}=1.2 (r_{\rm d}/R)^{1/2}P^{-1/2}$ deg for dipolar geometry.
Assuming that $\rho_{\rm d}=\rho_{\rm q}=\rho_{\rm obs}(W; \alpha, \beta)$
we get $r_{\rm d}^{1/2}/r_{\rm q}=s/R^{1/2}$, where $r_{\rm d}$ and $r_{\rm q}$
represent the emission altitudes for dipole and quadrupole field lines,
respectively.
Thus, if one allows nondipolar magnetic field in the emission region,
then  equation~(1) represents the upper limit for emission altitudes.
Moreover, the radius-to-frequency mapping would change from $r\propto
\nu^{-p}$ for dipolar field to $r\propto\nu^{-p/2}$
for quadrupole field, where the index $p$ is constrained
observationaly as $0.2\pm 0.1$ (see Table~4 in Paper II). 
Unfortunately, this constraint
is too weak to differentiate between dipole and quadrupole field lines.

Theoretically, a deviations from dipolar form can
be caused by (i) outflow of relativistic currents, (ii) field lines sweep-back
due to fast rotation, and (iii) currents flowing in the thin layer of the
neutron stars outer crust.
Let us briefly discuss these three cases:
\begin{description}
\item (i) The outflow of the charged component of the magnetospheric plasma
cannot change a structure of strong magnetic field generated inside the neutron
star. In fact, using the Ampere's law $\oint {\rm B}_{\rm t}\cdot 
{\rm ds}=(4\pi/c)J_{\rm p}$, 
where $B_{\rm t}$ is the toroidal component generated by the maximum 
available poloidal current through the polar cap 
$J_{\rm p}=4\pi c\rho_{\rm GJ}r_{\rm p}^2
\left(r_{\rm em}/R\right)^{3/2}$, 
where $r_{\rm em}\geq R$ is the emission altitude,
$r_{\rm p}\approx 10^4 P^{-1/2}$cm is the polar cap radius, 
$\rho_{\rm GJ}=B/cP$ is
the Goldreich-Julian (1969) corotational charge density, $c$ is the speed 
of light, $B$ is  a poloidal (dipolar) pulsar magnetic field and $P$ is the 
pulsar period. 
One can easily show that $B_t/B \approx 10^{-6}P^{-3/2}
\left(r_{\rm em}/R\right)^{3/2}$.
Thus, for typical pulsar $B_t$ is absolutely negligible,  while for the 
millisecond pulsars the toroidal contribution to  dipolar magnetic field 
can achieve a few percent at most.
\item (ii) The sweep-back of field lines by corotation with neutron star can 
occur near the light cylinder, where the corotation velocity approaches the
speed of light. However, it is generally accepted that the radio emission 
originates
at altitudes much smaller than the light cylinder radius (Cordes 1978, BCW,
Kramer et al. 1997).
\item (iii) The currents in the crust are most probably sources of multipolar 
components of the pulsar magnetic field outside the neutron star.
However, these contributions decay rapidly with increasing distance from the
star surface, and at radio emission altitudes only dipolar field can be 
expected, at least in normal pulsars. In the millisecond pulsars, in which
the emission region can be as close as few stellar radii from the polar cap, 
the magnetic field lines can, in principle, 
deviate significantly from dipolar form.
However, Arons (1993) has shown that the location of the 
spin-up line in the $P-\dot P$ diagram constrains the surface strength of 
multipole components to no more than 40\% of the dipole field in
millisecond pulsars.
\end{description}

\section{Conclusions}

The main conclusions from this paper and from previous work (Papers I and II)
on emission regions  in radio pulsars are  following:

\begin{description}
\item (1) A radius-to-frequency mapping operates in the emission region
and has a form of $r_{\rm em}(\nu) \propto \nu^{-0.21\pm 0.07}$.
  
\item (2) 
Pulsar radio emission is typically generated at 
altitudes smaller than a few per cent of the light-cylinder radius $r_{\rm LC}$
and the ratio $r_{\rm KG}/r_{\rm LC}$ decreases with increasing period.
The total extent of the emission region is smaller than
$250R$  in longer-period pulsars and correspondingly less in
shorter-period pulsars. 

\item (3) The emission region in old pulsars is located at correspondingly
lower altitudes than in young pulsars with approximately the same period.
\end{description}

In summary, pulsar emission altitudes depend on frequency $\nu$,
pulsar period $P$ and  pulsar age $\tau$, as 
described by equation~(1).

\section*{Acknowledgements}
 I gratefully acknowledge several helpful discussions
with Prof. J.~Gil. I~also thank the director of the Max-Planck
Institute for Radioastronomy  Prof. Dr. R.~Wielebinski 
for invitation and support. 
This work was in part supported by the Polish State Committee for Scientific
Reaserch Grant 2~P03D 008 19.

%\newpage

\section*{References}
\begin{description}
\item Arons J., 1993, ApJ, 408, 160 
\item Arzoumanian Z., Cordes J. M., Wasserman I., 1999, ApJ, 520, 696 (ACW)
\item Arzoumanian Z., Phillips J. A., Taylor J. H., Wolszczan A., 1996,
ApJ, 470, 1111
\item Bhattacharya D., 1996, in Johnston S., Walker M. A., Bailes M. eds., 
Proc. IAU Colloq. 160, Pulsars: Problems and Progress, Sydney,
A.S.P. Conference Series
\item Bell J. F., 1998, in Buccheri R., van Paradijs J., Alpar M. A.
eds., Proc. of the NATO ASI, The Many Faces of Neutron Stars, p. 159,
Kluwer Academic Publishers 
\item Blaskiewicz M., Cordes J. M., Wasserman I., 1991, ApJ,
370, 643 (BCW)
\item Camilo F., Thorsett S. E., Kulkarni S. R., 1994, ApJ, 421, L15
\item Cordes J. M., 1978, ApJ, 222, 1006
\item Cordes J. M., 1992, in Hankins T. H., 
Rankin J. M., Gil J. eds,
Proc. IAU Colloq. 128, The Magnetospheric Structure and Emission 
Mechanisms of Radio Pulsars, Univ. Zielona G\'{o}ra Press,
\L ag\'{o}w, Poland, p. 253
\item Cordes J. M., 1993, in van Riper K.A., Epstein R., Ho C. eds.,
Proc. of the Los Alamos Workshop, New Mexico,
Cambridge Univ. Press
\item Cordes J. M., Chernoff D. F., 1998, ApJ, 505, 315
\item Hoensbroech A.v., Xilouris K. M., 1997, A\&A, 324, 981 
\item Gil J., Gronkowski P., Rudnicki W., 1984, A\&A, 132, 312
\item Gil J.,  Kijak J., 1993, A\&A, 273, 563
\item Gil J., Sendyk M., 2000, ApJ, 541, 351 
\item Goldreich P., Julian W. H., 1969, ApJ, 157, 869
\item Gould M., 1994, PhD thesis, University of Manchester (G94)
\item Gould M., Lyne A. G., 1997, MNRAS, 301, 235
\item Gupta Y., Bhat R. N. D., Rao P. A., 1999, ApJ, 520, 173
\item Johnston S., Nicastro L., Koribalski B., 1998, MNRAS, 297, 108
\item Kijak J., Gil J., 1997, MNRAS, 288, 631 (Paper I)
\item Kijak J., Gil J., 1998, MNRAS, 299, 855 (Paper II)
\item Kramer M., Xilouris K. M., Jessner A., Lorimer D. R., Wielebinski R.,
Lyne A. G., 1997, A\&A, 322, 846
\item Kramer M., Xilouris K. M., Lorimer D. R., Doroshenko O.,
Jessner A., Wielebinski R., Wolszczan A., Camilo F., 1998, ApJ, 501, 270
\item Kramer M., Lange C., Lorimer D.R., Backer D. C., Xilouris K. M.,
Jessner A., Wielebinski R., 1999, ApJ, 526, 957
\item Lyne A. G., 1997, A\&A, 322, 846 
\item Lyne A. G., Anderson B., Salter M. J., 1982, MNRAS, 201, 503
\item Lyne A. G., Manchester R. N., 1988, MNRAS,  234, 477 (LM88)
\item Lyne A. G., Smith F., 1990, {\it Pulsar Astronomy}, Cambridge Univ.
Press
\item Lorimer D. R., 1996, in Johnston S., Walker M. A., Bailes M. eds., 
Proc. IAU Colloq. 160, Pulsars: Problems and Progress, Sydney,
A.S.P. Conference Series
\item Manchester R. N., Taylor J. H., 1977, {\it Pulsars}, San Francisco, 
Freeman
\item Phillips J. A., 1992, ApJ, 355, 282 
\item Rankin J. M., 1993, ApJ, 405, 285 (R93)
\item Thorsett S. E., Stinebring D. R., 1990, ApJ, 361, 644
\item Young M. D., Manchester R. N., Johnston S., 1999, Nature, 400, 848
\end{description}

\end{document}